\documentclass[twocolumn]{aastex631}

\begin{document}

\title{Precision Measurement of Large Shear Signals}

\author[0000-0001-7169-4642]{Jiarui Sun}
\affiliation{State Key Laboratory of Dark Matter Physics, School of Physics and Astronomy, \\
Shanghai Jiao Tong University, Shanghai 200240, China}

\author[0000-0003-0002-630X]{Jun Zhang}
\affiliation{State Key Laboratory of Dark Matter Physics, School of Physics and Astronomy, \\
Shanghai Jiao Tong University, Shanghai 200240, China}

\correspondingauthor{Jun Zhang}
\email{betajzhang@sjtu.edu.cn}

%\collaboration{20}{(AAS Journals Data Editors)}

\author[0009-0007-6825-4890]{Li Cui}
\affiliation{State Key Laboratory of Dark Matter Physics, School of Physics and Astronomy, \\
Shanghai Jiao Tong University, Shanghai 200240, China}

\author[0000-0002-6061-5977]{Alessandro Sonnenfeld}
\affiliation{State Key Laboratory of Dark Matter Physics, School of Physics and Astronomy, \\
Shanghai Jiao Tong University, Shanghai 200240, China}

\author[0000-0002-9373-3865]{Xin Wang}
\affiliation{School of Astronomy and Space Science, University of Chinese Academy of Sciences (UCAS), Beijing 100049, China}
\affiliation{National Astronomical Observatories, Chinese Academy of Sciences, Beijing 100101, China}
\affiliation{Institute for Frontiers in Astronomy and Astrophysics, Beijing Normal University, Beijing 102206, China}



\begin{abstract}
So far, estimators of galaxy shape distortions are only carefully studied perturbatively in the case of small shear signals, mainly for weak lensing science. However, in the neighborhood of massive foreground clusters, a large number of background galaxies can be significantly distorted. The measurement of such large shear signals could be quite nontrivial under general observing conditions, i.e., in the presence of the point spread function (PSF) and noise. In this work, we propose a non-perturbative method to exactly recover large shear signals ($\gtrsim 0.5$) under general conditions.  
We test the method on simulated galaxy images, and find that it is accurate down to the very faint end. This new method is particularly useful for more accurate recovery of the shear distribution in the neighborhood of massive foreground clusters, thereby improving the modeling of the underlying dark matter halo properties.
\end{abstract}



\section{Introduction} \label{sec:intro}

Gravitational lensing is a direct probe of the foreground density fluctuation \citep{1992ApJ...388..272K}. On large scales, lensing typically causes systematic shape distortion (in terms of ellipticity change) of background galaxies at the level of one percent, the statistics of which are often used to constrain the underlying cosmological model \citep{2022PhRvD.105b3515S,2023PhRvD.108l3518L, 2024SCPMA..6770413L, 2025arXiv250319441W}. On the other hand, on the scales of galaxy groups or clusters, the lensing distortion can be quite large, leading to multiple background images and giant arcs, which have become unique and effective probes of the foreground density profiles of groups/clusters \citep{2003ApJ...598..804K,2014ApJ...795..163U,2024MNRAS.529..802H, 2025arXiv250707629S}.  

A key technical issue in lensing studies is about accurate measurement of the background galaxy shapes. So far, development of shape/shear measurement methods is mainly discussed in the field of weak lensing. Given that weak lensing signals are very small, the focus of this area is mainly on overcoming various observational effects (PSF, noise, galaxy modelling, etc.), which can be larger than the shear signal by an order of magnitude or more \citep{2002AJ....123..583B,2008A&A...484...67P,2012MNRAS.425.1951R,2013MNRAS.434.1604Z,2014MNRAS.441.2528K,2018PASJ...70S..25M, 2021A&A...645A.105G, 2023MNRAS.520.2328Z, 2017arXiv170202600H, 2019ApJ...875...48Z}. In doing so, it is commonly assumed that the shear signal is small ($\sim 0.01$),  so that the observed galaxy ellipticity can be approximated as a linear function of the underlying shear. 
However, this approximation may break down at the vicinity of the foreground cluster, where strong shear signals ($g \gtrsim 0.5$) can induce nonlinear effects.  
Current shear measurement methods in this case could lead to significant biases. For example, \citet{2020A&A...640A.117H} employed the KSB+ formalism \citep{1995ApJ...449..460K, 1997ApJ...475...20L, 1998ApJ...504..636H} to investigate the impact of strong shear. Their results demonstrate that certain measurements notably deviate from the linear trend under strong shear conditions. 
Shear measurements based on model fitting can be unaffected by the nonlinear effects if the galaxy shape is perfectly elliptical, as we briefly discuss in the end of \S\ref{FQN_method}. In practice, however, simple models may suffer from significant model biases \citep{2013MNRAS.434.1604Z, 2014ApJS..212....5M, 2014MNRAS.441.2528K, 2024A&A...691A.319E}. 
The purpose of this work is to introduce an unbiased way of estimating large shear signals under general observing conditions. 

Our new method is closely related to the Fourier\_Quad method (\citealt{2015JCAP...01..024Z, 1997ApJ...475...20L}, FQ hereafter) , but with significant differences. In the FQ method, the shear estimators are simple functions of the multipole moments of the galaxy image's power spectrum. It has been shown explicitly that the method is accurate to the second order in shear \citep{2011JCAP...11..041Z}, without assumptions on the morphologies of the galaxy or the PSF. Furthermore, its robustness for cosmic shear measurements has been validated using real data, showing reliable performance even for poorly resolved images \citep{2019ApJ...875...48Z, 2022AJ....164..128Z}. 
Nevertheless, we find it extremely difficult to achieve higher order accuracy by extending the current formalism. 
Our new method gives up the idea of requiring the shear estimators to have specific functional forms. Instead, we find that one can introduce pseudo shear components to distort the coordinates in the definitions of the image moments in Fourier space, and vary their values to nullify the quadrupole moments. It is straightforward to show that the resulting pseudo shears are unbiased estimators of the underlying true shear signals, regardless of their amplitudes.   

In \S\ref{sec:style}, we present the details of the new method. In \S\ref{sec:floats}, we test the performance of our new method using numerical simulations, with different morphological forms of the galaxy and the PSF, and different levels of noise. We summarize our findings in \S\ref{sec:cite}.

\section{The Method} \label{sec:style}

\subsection{The Fourier\_Quad Method}
 
Let us define the intrinsic galaxy surface brightness distribution as $f_{S}\left(\vec{x}^{S}\right) $, and the lensed galaxy as $f_{L}\left(\vec{x}^{L}\right)$. 
The transformation between the unlensed and lensed coordinates is $\vec{x}^{S}  =  {\mathrm M} \vec{x}^{L} $, where ${\mathrm M}$ is the lensing distortion matrix defined as: 
\begin{equation}
{\mathrm M} = (1-\kappa)\left(\begin{array}{cc}
1-g_{1} & -g_{2} \\
-g_{2} & 1+g_{1}
\end{array}\right) \, ,
\label{eq:shear-matrix}
\end{equation}
where $\kappa$ is the convergence, describing the change in galaxy size due to lensing, and $g$ is the reduced shear, causing the stretching of the images. 
The lensed image should be further convolved by the PSF to form the observed image, which can be written as: 
\begin{equation}
    f_{O}\left(\vec{x}^{O}\right) = \int d^{2} \vec{x}^{L} W_{P S F}\left(\vec{x}^{O}-\vec{x}^{L}\right) f_{L}\left(\vec{x}^{L}\right) \, ,
\end{equation}
where $W_{P S F}$ represents the form of the PSF. 
In the Fourier space, we have: 
\begin{equation}
  \tilde{f}_{O}(\vec{k})  =  \tilde{W}_{P S F}(\vec{k}) \tilde{f}_{L}(\vec{k}) \, .  
\end{equation}
It can be shown that the reduced shear can be estimated using the multipole moments of the galaxy image's power spectrum \citep{2011JCAP...11..041Z}:

\begin{eqnarray}
&&\frac{1}{2} \frac{\left\langle P_{20}-P_{02}\right\rangle}{\left\langle P_{20}+P_{02}-\beta^{2} D_{4} / 2\right\rangle}  =-g_1+O\left(g^3\right)  , \\ \nonumber
&&\frac{\left\langle P_{11}\right\rangle}{\left\langle P_{20}+P_{02}-\beta^{2} D_{4} / 2\right\rangle}  =-g_2+O\left(g^{3}\right) \, ,
\end{eqnarray}
where
\begin{eqnarray}
&&P_{i j} =\int d^{2} \vec{k} k_{1}^{i} k_{2}^{j} T(\vec{k}) \left|\widetilde{f}_{O}(\vec{k})\right|^{2} \, , \\ \nonumber
&&D_{n} =\int d^{2} \vec{k}|\vec{k}|^{n} T(\vec{k}) \left|\widetilde{f}_{O}(\vec{k})\right|^{2} \, , \\ \nonumber
&&T(\vec{k})=|\tilde{W}_{\beta}(\vec{k})|^2/|\tilde{W}_{P S F}(\vec{k})|^2 \, .
\end{eqnarray}
$\tilde{W}_{\beta}$ ($=\exp (-\beta^{2} k^{2} / 2)$) is the power of the isotropic Gaussian PSF, and $\beta$ is its scale radius. 
The factor $T(\vec{k})$ is applied to transform the PSF into the isotropic Gaussian form\footnote{Generalization to elliptical Gaussian forms is possible, but would result in modified formulations. This topic is however beyond the scope of this work. } , standardizing the PSF model for computational tractability while maintaining analytical simplicity. 
This method is only accurate to the second order in shear. We aim to develop an approach that extends it to higher orders. We find it hard to achieve this goal by using higher order moments in FQ. 
Alternatively, we consider nullifying the quadrupole moments by distorting the coordinates in the definitions 
of moments using pseudo shear signals. We therefore refer to it as Fourier\_Quad\_Nulling (FQN). This turns out to be a viable way, as we show in the next section.

\subsection{The FQN Method}
\label{FQN_method}
We introduce a new signal $g'$ to alter the kernel in the definition of the quadrupole moment: 
\begin{equation}
    P_{i j}'=\int d^{2} \vec{k}\left({\mathrm M}' \vec{k}\right)_{1}^{i}\left({\mathrm M}' \vec{k}\right)_{2}^{j} \frac{|\tilde{W}_{\beta}({\mathrm M}' \vec{k})|^2}{|\tilde{W}_{P S F}(\vec{k})|^2} \left|\tilde{f}_{
O}(\vec{k})\right|^{2}  \, , 
\label{eq:multiple-moment}
\end{equation}
where ${\mathrm M}^{\prime}$ is a transformation matrix, which is defined as  
\begin{equation}
{\mathrm M}' = \left(\begin{array}{cc}
1-g_{1}' & -g_{2}' \\
-g_{2}' & 1+g_{1}'
\end{array}\right) \, . 
\label{eq:alter-matrix}
\end{equation}
According to the definition, we can see that this operation is somewhat equivalent to altering the pre-seeing image of the galaxy, but without actually de-convolving the PSF.   
Note that to avoid numerical instability, the FWHM (Full Width at Half Maximum) of the target PSF ($W_{\beta}$) must be somewhat larger than that of the original PSF ($W_{PSF}$). 

We search for the best values of $g_{1,2}'$ to nullify the complex quadrupole moments: $P_{20}'-P_{02}'+i\left(2 P_{11}'\right)$. It can be shown that the resulting $g_{1,2}'$ are related to the intrinsic galaxy shape $g_{1,2}^S$ and the shear signal $g_{1,2}$ through a simple relation in the complex form (i.e., $g=g_1+ig_2$, and similarly for $g^S$ and $g'$ ):
\begin{equation}
    g' = -\frac{g^S + g}{1 + g^S \cdot g^*} +\mathrm{noise} \, .
\label{eq:gpp-gp-g}
\end{equation}
where $g^*$ is the complex conjugate of $g$. 
By averaging $g'$, we can eliminate the shape noise contribution due to the fact that $\left\langle (g^S)^n \right\rangle=0$ for any integer value of n. Finally, we get:
\begin{equation}
    \left\langle g' \right\rangle = -g \, .
\end{equation}
The detailed proof is provided in the appendix.

Note that a relation similar to Eq.(\ref{eq:gpp-gp-g}) exists between the shear and the galaxy ellipticity (without PSF) if the ellipticity is defined as $\epsilon=(1-q)/(1+q) e^{2 i \phi}$, in which $q$ is the minor-to-major axis ratio, and $\phi$ is the position angle. The definition of $\epsilon$ however relies on the assumption that the galaxy morphology is a perfect ellipse, which is not quite realistic in practice. 

Numerically, we determine $g'$ by first establishing a prior from other measurement (e.g., from FQ), and then search for the value of $g'$ in the neighborhood of this prior so that the amplitudes of the quadrupole moments reach their minimum values. The technical details of the $g'$ measurement are provided in \S\ref{sec:floats}. 

\section{The Numerical Test} \label{sec:floats}

We demonstrate the reliability of the new method with numerical experiments. We simulate both regular and irregular galaxy morphologies utilizing point sources with realistic brightness distributions. Each galaxy simulation is placed on a $128 \times 128$ stamp. 
For regular galaxies, we uniformly distribute points within a circle of a fixed radius The flux of each point depends on its distance from the center, so that the overall profile follows a Sérsic profile \citep{1968adga.book.....S}, which is defined as: 
\begin{equation}
    I(R)=I_e\exp\left\{-b_n\left[\left({R\over{R_e}}\right)^{1/n}-1\right]\right\} \, .
\label{eq:sersic}
\end{equation}
The parameter $n$ is the Sérsic index, $R_e$ is the projected half-light radius, $I_e$ is the
intensity at $R_e$, and $b_n = 1.999n - 0.327$ for $n \ge 1$ \citep{1989woga.conf..208C}. We assign the ellipticity to the galaxies using matrix transformations. 
Irregular galaxies are simulated by the 2D random walks to generate the positions of the point sources, following the procedures described in \citet{2008MNRAS.383..113Z}. The number of steps is 1000, with a threshold distance of 8 pixels. All length measurements hereafter are expressed in unit of the pixel size. 

Given the positions and brightness of these discrete points, we perform a convolution operation  
with two kinds of PSF models, the Gaussian PSF and the Moffat PSF, given by:
\begin{equation}
\begin{array}{cc}
W_{Guass}(x,y) \propto \exp [-(x^2+y^2)/(2\sigma^2)] \\
W_{Moffat}(x,y) \propto [1+(x^2+y^2)/r^2]^{-b} \, .
\label{eq:psf-model}
\end{array}
\end{equation}

\subsection{Measurement without Noise} \label{subsec:mock-regular}

Differing from the FQ method, the new measurement entails finding $g'$ to drive the new quadrupole moments towards zero. The process can be simply described as follows. 
We adopt the results from FQ as a prior and sample points around this value to generate initial guesses. To determine the minimum, we first fix $g_2^\prime$ and fit the relationship between $g_1^\prime$ and the absolute quadrupole moment $|P_{20}^\prime - P_{02}^\prime + i(2P_{11}^\prime)|$. Then we fix $g_1^\prime$ and optimize for $g_2^\prime$. This iterative process continues until the difference between successive values of $g_{1,2}^\prime$ falls below $5 \times 10^{-4}$. The converged solution yields the final $g^\prime$. 
 
\begin{figure}
    \centering
    \includegraphics[width=0.45\textwidth]{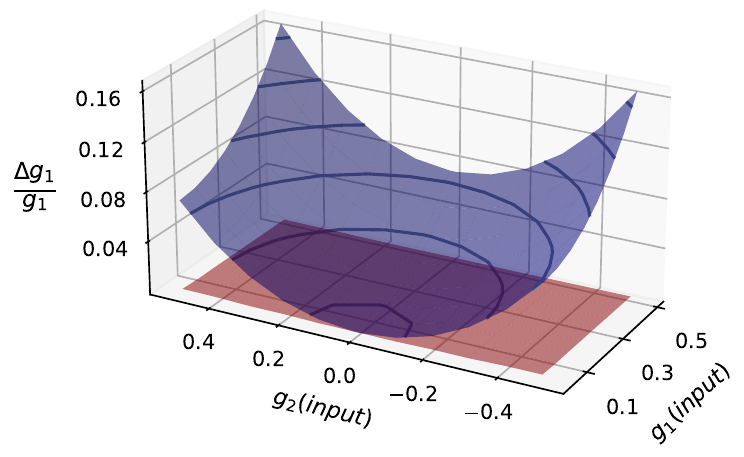}
    \caption{An example showing the behavior of the shear bias when the shear signals become large. The blue and red surfaces refer to the results of the FQ and FQN method respectively. The contours correspond to the values of 0.005, 0.035, 0.065, 0.095, 0.125 and 0.155 respectively. }
    \label{fig:moffat-nonoise}
\end{figure}

We use four identical galaxies, generated by rotating a single galaxy by 45 degrees for four times, to suppress the shape noise. 
Fig.\ref{fig:moffat-nonoise} shows one of the examples for the relative errors. The simulation adopts the regular galaxies, with the shape parameters as: $R_e=4$, $n=2$, and $|g^S|=0.2$. We use the Moffat-profile for PSF models with $r=4$ and $b=3$. Here, the x- and y-axes correspond to the input shear values, while the z-axis represents the relative errors between measured and expected signals. The FQ method provides accurate results (shown as the blue surface) for small shear regimes $(g \lesssim 0.1)$, but discrepancies emerge as the shear signal increases. Notably, the FQN measurement results (shown as the red surface) remain accurate across all tested magnitudes. We only show the results for $g_1$, because the results for $g_2$ are similar.  

We have tested the accuracy of FQN for irregular galaxies as well, and the conclusions are similar. We also change the other conditions in our tests, including the PSF models, and relative scales between the PSF and galaxy, etc.. In all cases, we find that the FQN method yields negligible bias when the shear signal becomes large, consistent with the results of Fig.\ref{fig:moffat-nonoise}.

\subsection{Measurement with Noise}
\label{subsec:mock-noise}

Photon noise changes the shape of the galaxy, and potentially introduces systematic bias in shape/shear measurement.  
For convenience, we add Gaussian noise to the simulation to further test the reliability of FQN in the presence of noise.

When noise is present, the $\left|\tilde{f}_{
O}(\vec{k})\right|^{2}$ term in Eq.(\ref{eq:multiple-moment}) needs to be replaced with:
\begin{equation}
\label{ff}
    \left|\tilde{f}_{
O}(\vec{k})\right|^{2}-F^O-\left|\tilde{f}_{
B}(\vec{k})\right|^{2}+F^B  \, , 
\end{equation}
where $f_B$ is the image of background noise, and $F^O$ and $F^B$ represent the Poisson noise power spectra on the source and background images respectively \citep{2015JCAP...01..024Z}.
It represents the power spectrum of the galaxy image after background correction and Poisson noise subtraction. 
For SNR$\, \gtrsim 50$, we find that accurate shear measurements can be obtained through individual galaxy analysis followed by ensemble averaging. However, when the SNR is low, noise induces noticeable fluctuations in quadrupole moments, making it hard to find the minimum value. In this case, we can instead nullify the average of the quadrupole moments of a number of galaxies with the same pseudo shear signals to mitigate the impact of noise-induced disturbances on the results. 
For combining the quadrupole moments, one can apply different weightings to galaxies of different SNRs and ellipticities. For the simplified scenario in this section, we proceed without weighting, and plan to discuss the weighting scheme in another work when applying our method to real data. 
We construct a two-dimensional surface by evaluating the absolute values of the summed quadrupole moments as a function of the pseudo shear components $(g_1',g_2')$. The final shear estimator $g'$ is determined by locating the minimum of the fitted 2D surface. 

\begin{figure}[t]
    \centering
    \includegraphics[width=0.45\textwidth]{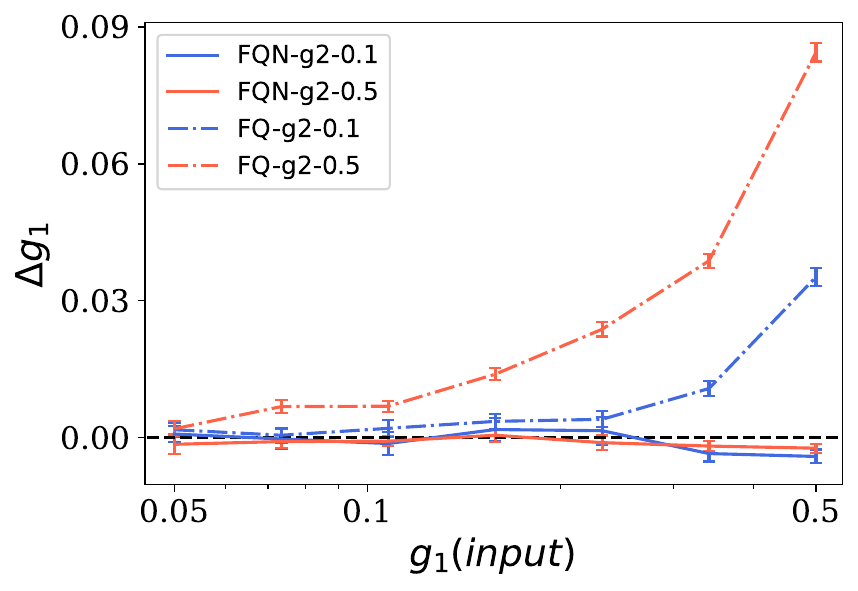}   \includegraphics[width=0.45\textwidth]{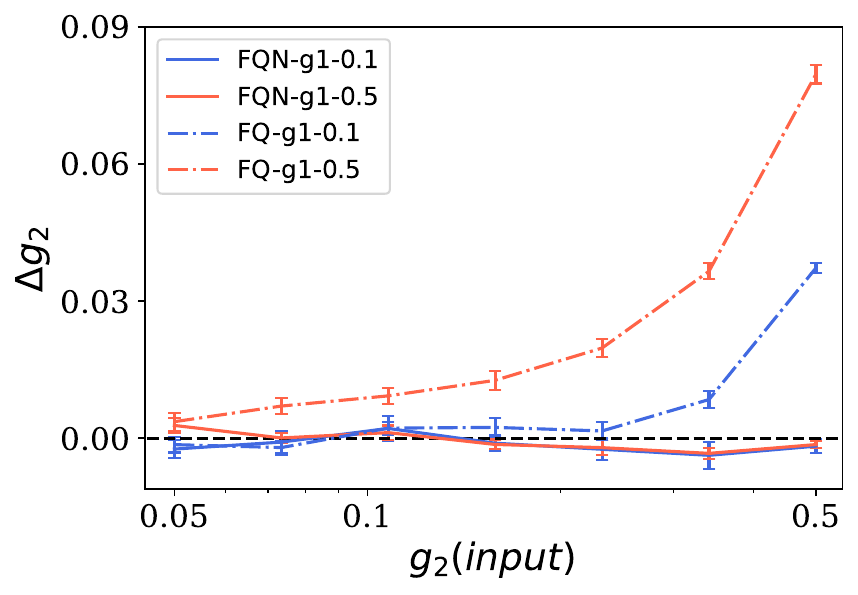}
    \caption{An example of the shear errors in regular galaxies at SNR = 40.  Different colors in the plots represent different $g_2$ ($g_1$) input values. The dot-dashed curves correspond to the FQ results, while the solid lines correspond to the FQN results. The black dashed line indicates the zero baseline for comparison. 
    }
    \label{fig:moffat-noise}
\end{figure}

\begin{figure}[h]
    \centering
    \includegraphics[width=0.45\textwidth]{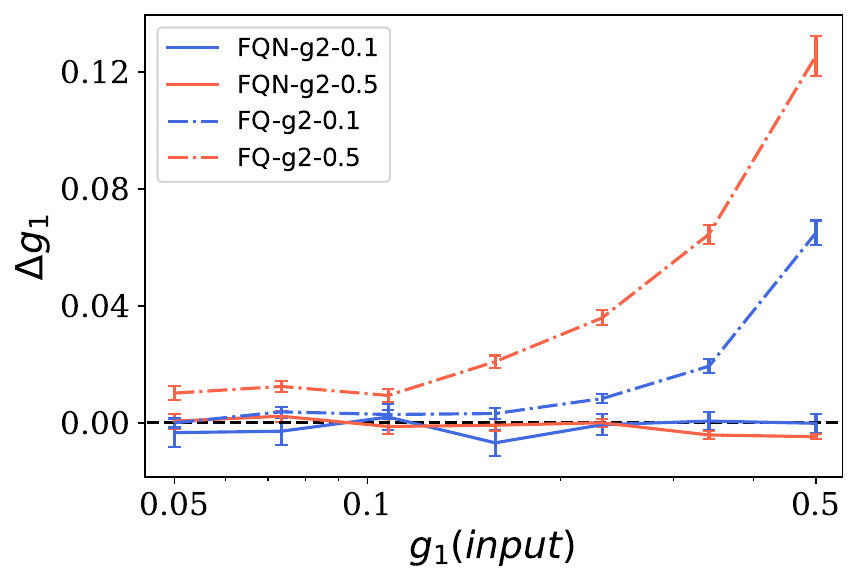}   
    \caption{An example of the shear errors using irregular galaxies with SNR = 10. The color scheme and line styles are the same as those in Fig.\ref{fig:moffat-noise}. 
    }
    \label{fig:moffat-noise-snr10}
\end{figure}

An example of the results is shown in Fig.\ref{fig:moffat-noise}, where we choose the galaxy images with SNR=$40$. 
For each shear component, we simulate 10,000 randomly oriented galaxies\footnote{We utilize this large galaxy sample to conduct a precision test. However, achieving such high galaxy number counts is unrealistic with current observational capabilities. For instance, even in dense fields such as Abell 2744 observed by the JWST (James Webb Space Telescope), the galaxy number density only reaches approximately 200 galaxies per arcmin$^2$ \citep{2024MNRAS.529..802H, 2024ApJ...961..186C}. }.  
Selection of other parameters is consistent with Fig.\ref{fig:moffat-nonoise}. The upper plot displays the deviation for different $g_1$ with a fixed $g_2$, and the lower plot shows the deviation of $g_2$ with a fixed $g_1$. The x-axis represents the input signal and the y-axis represents the absolute errors. The error bars are calculated using the Jackknife method with a total of forty groups. It can be observed that for the FQ results, the bias increases with the signal magnitude, whereas the results for FQN remain accurate. An alternative test demonstrates that even with SNR$=10$, the FQN method could still yield reliable results. The results are shown in Fig.\ref{fig:moffat-noise-snr10} for $g_1$. In this test, we use a sample of 100,000 irregular galaxies. 

\section{Conclusion and Discussion} \label{sec:cite}

Existing shear measurement methods are usually studied in the field of weak lensing, implying the smallness of the shear signals. In this paper, based on the FQ method, we propose a new method to accurately recover large shear signals under general observing conditions. The new method is called Fourier\_Quad\_Nulling, because it adds pseudo shear signals to nullify the quadrupole moments in Fourier space. 
When the quadrupole moments reach zero, the pseudo shears are unbiased estimators of the underlying shear signal, regardless of their amplitudes. We test our method on both regular and irregular galaxies in numerical simulations, with different PSF models and noise levels. The results demonstrate the robustness of our method under general conditions.

In practice, we can think of two ways for shear recovery: 1. for galaxies with high SNR, measurement on individual galaxy image can be obtained, and later used for different types of shear statistics; 2. for low SNR images, a joint analysis of multiple galaxies is required, as shown in \S\ref{subsec:mock-noise}. In real data, the second type of measurement can be used for the statistics of shear stacking in galaxy-galaxy lensing or clustering lensing. Another scenario is about making shear map around a massive foreground cluster, in which we can model the local shear field with a small number of parameters, and find their values by nullifying the quadrupole moments of the galaxy images in the neighborhood altogether.

We caution that for galaxies with low SNR, reliable shear estimators in FQN cannot be extracted from individual galaxy images, resulting in high computational complexity in shear statistics. For instance, in galaxy-galaxy lensing, traditional methods can directly compute the tangential shear by rotating shear estimators, whereas our FQN method requires rotating the galaxy images and re-finding the solutions. Furthermore, since cosmic shear signals are typically weak ($\lesssim 0.01$), conventional methods already provide reliable measurements. FQN is particularly suited for recovering the shear distribution in the neighborhood of massive galaxy clusters, thereby placing tighter and more accurate constraints on the properties of the underlying halo density profiles. This is what we aim to do in a separate work.

%% Also note that the akcnowlodgment environment does not support long amounts of text. If you have a lot of people and institutions to acknowledge, do not use this command. Instead, create a new \section{Acknowledgments}.

\begin{acknowledgments}
This work is supported by the National Key Basic Research and Development Program of China (2023YFA1607800, 2023YFA1607802), 
and the science research grants from China Manned Space Project (No. CMS\-CSST\-2021\-A01). XW is supported by the National Natural Science Foundation of China (grant 12373009), the CAS Project for Young Scientists in Basic Research Grant No. YSBR-062, the Fundamental Research Funds for the Central Universities, the Xiaomi Young Talents Program, and the China Manned Space Program with grant no. CMS-CSST-2025-A06. The computations in this paper were run on the $\pi$ 2.0 cluster supported by the Center of High Performance Computing at Shanghai Jiaotong University.
\end{acknowledgments}

\appendix
\section{Detailed Derivation of FQN}
\label{sec:prove1}

Suppose the surface brightness distribution $f_{S}(\vec{x}^{S})$ of the
galaxy can be linearly modified to become $f_C(\vec{x}^C)$, so that its quadrupole moments disappear, i.e., $P_{20}^C-P_{02}^C=P_{11}^C=0$, in which: 
\begin{equation}
\label{pc}
P_{i j}^{C}=\int d^{2} \vec{k} \vec{k}_{1}^{i} \vec{k}_{2}^{j}\left|\widetilde{W}_{\beta}\left(\vec{k}\right) \tilde{f}_{
C}(\vec{k})\right|^{2}\, .
\end{equation}
To do so, we define $ \vec{x}^{C}  =  {\mathrm M}^{S} \vec{x}^{S}$ , $f_{S}\left(\vec{x}^{S}\right)  =  f_{C}\left(\vec{x}^{C}\right)$, and ${\mathrm M}^{S}$ is a transformation matrix, defined as 
\begin{equation}
{\mathrm M}^{S} = \left(\begin{array}{cc}
1-g_{1}^S & -g_{2}^S \\
-g_{2}^S & 1+g_{1}^S
\end{array}\right) \, .
\label{eq:circle-Matrix}
\end{equation}
We can call $g_{1,2}^S$ the intrinsic shape/ellipticity of the galaxy.
The surface brightness field of the lensed galaxy and its circularized image in Fourier space are given by: 
\begin{equation}
\tilde{f}_{L}(\vec{k}) = \left|\mathrm{det}\left( {{\mathrm M}^{-1}}^T\right)\right| \tilde{f}_{S}\left({{\mathrm M}^{-1}}^T \vec{k}\right)   
= \left|{\mathrm{det}}\left({\left( 
 {\mathrm M}{\mathrm M}^{S} \right)  }^{-1}\right)\right| \tilde{f}_{C}\left( {\left( 
{\mathrm M} {\mathrm M}^{S} \right) }^{-1} \vec{k}\right) \, .  
\end{equation}
With the additional kernel ${\mathrm M}'$, the modified multipole moments can be expressed as: 
\begin{eqnarray}
P_{i j}^{\prime}&=&\int d^{2} \vec{k}\left({\mathrm M}^{\prime} \vec{k}\right)_{1}^{i}\left({\mathrm M}^{\prime} \vec{k}\right)_{2}^{j} \frac{|\tilde{W}_{\beta}({\mathrm M}^{\prime} \vec{k})|^2}{|\tilde{W}_{P S F}(\vec{k})|^2} \left|\tilde{f}_{
O}(\vec{k})\right|^{2} \nonumber \\ \nonumber
&=& \left|{\mathrm{det}}\left({\left( 
 {\mathrm M}{\mathrm M}^{S} \right)  }^{-1}\right)\right|^2 \int d^{2} \vec{k}\left({\mathrm M}^{\prime} \vec{k}\right)_{1}^{i}\left({\mathrm M}^{\prime } \vec{k}\right)_{2}^{j} \times \left|\widetilde{W}_{\beta}\left({\mathrm M}^{\prime}\vec{k}\right) \tilde{f}_{C}({\left( 
{\mathrm M} {\mathrm M}^{S} \right) }^{-1}\vec{k})\right|^{2} \nonumber \\ 
\label{eq:Pij-3}
&=&\left|\mathrm{det}\left(\left({\mathrm M} {\mathrm M}^{S}\right)^{-1}\right)\right| \int d^{2} \vec{k}\left({\mathrm M}^{\prime} {\mathrm M} {\mathrm M}^{S} \vec{k}\right)_{1}^{i} \left({\mathrm M}^{\prime} {\mathrm M} {\mathrm M}^{S} \vec{k}\right)_{2}^{j}\left|\widetilde{W}_{\beta}\left({\mathrm M}^{\prime} {\mathrm M} {\mathrm M}^{S} \vec{k}\right) \tilde{f}_{C}(\vec{k})\right|^{2} \, . 
\end{eqnarray}
The matrix multiplication can be done directly as:
\begin{equation}
{\mathrm M} {\mathrm M}^{S}=(1-\kappa) \left(\begin{array}{cc}
1-g_1 & -g_2 \\
-g_2 & 1+g_1
\end{array}\right) \left(\begin{array}{cc}
1-g_1^S & -g_2^S \\
-g_2^S & 1+g_1^S
\end{array}\right) 
=(1-\kappa)\left(1- \hat{\kappa} \right)\left(\begin{array}{cc}
1-\hat{g}_1 & -\hat{g}_2+\delta \\
-\hat{g}_2-\delta & 1+\hat{g}_1
\end{array}\right)  \, ,    
\end{equation}
in which $\hat{\kappa} = -(g_1^S g_1 + g_2^S g_2)$ , $\hat{g}_i = (g_i^S + g_i)/(1 - \hat{\kappa})$, $\delta = (-g_1^S g_2 + g_2^S g_1)/(1 - \hat{\kappa})$. Furthermore, we have:
\begin{equation}
\begin{array}{cc}
{\mathrm M}^{\prime} {\mathrm M} {\mathrm M}^{S} &= (1-\kappa) (1-\hat{\kappa}) \left(\begin{array}{cc}
1-g_1' & -g_2' \\
-g_2' & 1+g_1'
\end{array}\right) \left(\begin{array}{cc}
1-\hat{g}_1 & -\hat{g}_2+\delta \\
-\hat{g}_2-\delta & 1+\hat{g}_1
\end{array}\right) \\
&= (1-\kappa) (1-\hat{\kappa}) (1-\bar{\kappa})\left(\begin{array}{cc}
1-\bar{g}_1 & -\bar{g}_2+\bar{\delta} \\
-\bar{g}_2-\bar{\delta} & 1+\bar{g}_1
\end{array}\right) \, ,
\end{array}
\label{eq:N}
\end{equation}
where 
$$
\bar{\kappa} = -(g_1' \hat{g}_1 + g_2' \hat{g}_2) \, , \bar{g}_1 = \frac{\hat{g}_1 + g_1' - g_2' \delta}{1 - \bar{\kappa}} \, , \bar{g}_2 = \frac{\hat{g}_2+ g_2' + g_1' \delta}{1 - \bar{\kappa}}\, , \bar{\delta} = \frac{\delta +g_1' \hat{g}_2 - g_2' \hat{g}_1}{1 - \bar{\kappa}} \, . 
$$ 

Substituting Eq.(\ref{eq:N}) into Eq.(\ref{eq:Pij-3}) and applying a Taylor expansion, the complex quadrupole moments can be expressed as:
\begin{equation}
\begin{array}{cc}
P_{20}^{\prime}-P_{02}^{\prime}+i\left(2 P_{11}^{\prime}\right) \\
=\left|{\mathrm{det}}\left({\left( 
 {\mathrm M}{\mathrm M}^{S} \right)  }^{-1}\right)\right| \int d^{2} \mathrm{k} \left\{(1-\tilde{\kappa})^{2}\left[(1-\mathrm{i} \bar{\delta})^{2} \mathrm{k}^{2}+\bar{g}^{2} \mathrm{k}^{* 2}-2 \bar{g}(1-i \bar{\delta})|\mathrm{k}|^{2}\right]\right\} \\
\cdot \exp \left(-\beta^{2}|\mathrm{k}|^{2}\right)\left(1+\Delta+\frac{1}{2} \Delta^{2}+\frac{1}{6} \Delta^{3} + \dots \right)\left|\tilde{f}_{C}(\mathrm{k})\right|^{2} \, , \label{eq:P20}
\end{array}
\end{equation}
where $\tilde{\kappa}=(1-\kappa) (1-\hat{\kappa}) (1-\bar{\kappa})-1$ , $\Delta=- \beta^2 \left\{ \left(-2 \tilde{\kappa}+\tilde{\kappa}^{2}\right)|\mathrm{k}|^{2}+(1-\tilde{\kappa})^{2}\left[\left(|\bar{g}|^{2}+\bar{\delta}^{2}\right)|\mathrm{k}|^{2}-\bar{g}^{*}(1-i \bar{\delta}) \mathrm{k}^{2}-\bar{g}(1+i \bar{\delta}) \mathrm{k}^{* 2}\right] \right\}$. Note that both k and $\bar{g}$ are complex numbers. The term proportional to $\Delta^n$ is: 
\begin{equation}
    \int d^{2} \mathrm{k}  \cdot \exp \left(-\beta^{2}|\mathrm{k}|^{2}\right) \left|\tilde{f}_{C}(\mathrm{k})\right|^{2} \cdot \left[(1-\mathrm{i} \bar{\delta})^{2} \mathrm{k}^{2}+\bar{g}^{2} \mathrm{k}^{* 2}-2 \bar{g}(1-i \bar{\delta})|\mathrm{k}|^{2}\right] \cdot \Delta^n  \, ,
\end{equation}
in which the only polynomial terms that do not contain $\bar{g}$ are $\mathrm{k}^2|\mathrm{k}|^{2n}$. The term with $n=0$ yields zero by the definition of $f_C$. The others with $n>0$ correspond to high order moments of $f_C$, which do not necessarily vanish because the ellipticities of an actual galaxy’s isophotes may vary with radius. They can however be regarded as noise. 

To sum up, $P_{20}^{\prime}-P_{02}^{\prime}+i\left(2 P_{11}^{\prime}\right)$ can be expressed as terms proportional to $\bar{g}$ plus noise terms resulting from high order moments of the galaxy shape. Therefore, when it approaches zero, $\bar{g}$ is only related to noise, leading to 
\begin{eqnarray}
&&g_{1}{ }^{\prime} =-\left(\hat{g{_{1}}}+\delta \hat{g{_{2}}}\right) /\left(1+\delta^{2}\right)+\mathrm{noise} \nonumber  \\ 
&&g_{2}{ }^{\prime} =-\left(\hat{g{_{2}}}-\delta \hat{g{_{1}}}\right) /\left(1+\delta^{2}\right)+\mathrm{noise}  \, .
\end{eqnarray}
Consequently, we have: 
\begin{equation}
    g^{\prime}=-\hat{g}/(1+i\delta)+\mathrm{noise}=-\frac{g^{S} + g}{1 + g^{S} \cdot g^*} +\mathrm{noise} \, .     
\end{equation}
Since the intrinsic ellipticity of the galaxy does not have a preferred orientation, we have $\left\langle (g^S)^n \right\rangle =0$ for all $n>0$, $n \in \mathbb{Z}$. We think it is also reasonable to assume that the shape noise caused by high order moments of the galaxy shape is statistically independent of the quadrupole moments. It can therefore be eliminated by averaging. 
Based on the above discussion, $g^{\prime}$ is an unbiased estimator of the reduced shear, i.e.,
$
    \left\langle g^{\prime} \right\rangle = -g \, .
$

\bibliography{sample631}{}
\bibliographystyle{aasjournal}



\end{document}